\documentclass[12pt]{article}
\begin{document}

\begin{center}{\LARGE\bf Remarks on DSR and Gravity}\end{center}

\vspace{3mm}

\begin{center}{\Large F. Hinterleitner}\end{center}
\begin{center}{\em Institute of Theoretical Physics and Astrophysics,\\
Faculty of Science of the Masaryk University,\\
Kotl\'{a}\v{r}sk\'{a} 2, 611 37 Brno,
Czech republic} \end{center}

\vspace{3mm}

\small{\bf Abstract.} Modifications of Special Relativity by the
introduction of an invariant energy and/or momentum level (so-called
Doubly Special Relativity theories, DSR) or by an energy-momentum
dependence of the Planck constant (Generalized Uncertainty
Principle, GUP) are compared with classical gravitational effects in
an interaction process. For the low energy limit of the usual
formulations of DSR to be equivalent to Newtonian gravity, a
restrictive condition is found. GUP yields an effective repulsion,
in analogy to gravitational repulsion in loop quantum cosmology.

\section{Introduction}
Tentative quantum theories of gravity -- string theory as well as
loop quantum gravity or non-commutative geometry -- indicate the
existence of an invariant length scale of the order of magnitude of
the Planck length, which is in obvious contradiction with standard
Lorentz symmetry, when taken seriously to arbitrary small scales.
For this reason in the last years various attempts were made to
modify Special Relativity (SR) \cite{GAC,MS} in such a way that one
(or more) invariant quantities in addition to the speed of light
would be reconciled with the relativity principle. Theories of this
kind are called Doubly or Deformed Special Relativity (DSR).

Early examples were formulated in momentum space by the application
of nonlinear representations of the Lorentz group to the energy and
momentum of physical objects, such that there is an invariant value
of energy and/or momentum of the order of the Planck energy and the
Planck momentum. Technically this may be achieved by splitting
energy/momentum variables into ``physical" ones, usually denoted by
$E$ and $\vec p$, and ``pseudo"-variables (sometimes called platonic
variables) $\varepsilon$ and $\vec\pi$, with both kinds of variables
related by an invertible nonlinear transformation. Pseudo-variables
satisfy the usual linear relations of SR, in consequence, the
physical ones are acted upon by the (boost sector of the)
Poincar\'{e} group in a deformed, nonlinear way. Note that the
denomination `deformed', common in the present context, denotes
merely the action of the Lorentz group and has nothing to do with
deformations of the Poincar\'{e} Lie algebra.

Pseudo-variables, although being mere auxiliary quantities in the
construction of modified Lorentz transformations, have the following
formal significance. Providing the linear representation of the
Poincar\'{e} group, they carry the usual vector space structure of
SR momentum space, whereas the space of $E$ and $\vec p$ becomes
curved. Therefore, when subsystems are composed to a whole, it is
the pseudo-variables which must be additive in order to preserve the
underlying Lorentz group structure. For this reason in the
calculations of reaction thresholds or cross sections of particle
interactions in the framework of DSR, conservation rules are
formulated in terms of them \cite{JV,LN}. This leads to a violation
of ordinary energy/momentum conservation; particularly the total
energy/momentum of a composite system never exceeds the invariant
values, as long as the energy/momentum of its components are below
them. (This is the so-called soccer-ball problem, see, e.\,g.
\cite{Sta,soc}.)

Possible physical consequence are anomalies of reaction thresholds
and an energy dependence of the speed of light in some versions of
DSR. Even if the effect is tiny (of the order {\sl photon
energy/Planck energy}), it might become measurable when photons run
over cosmic distances. A recently observed slight energy dependence
of the time of arrival of photons from a $\gamma$ ray flare might be
interpreted in this sense, if we knew the mechanism of emission, see
\cite{magic}.

More advanced versions of DSR are completed by modified space-time
Lorentz transformations, associated to the transformations in
momentum space in different ways. Some of these attempts assume
momentum space to be a de Sitter space \cite{sitt}, other ones make
use of Hopf algebra techniques \cite{hopf}. These approaches lead to
non-commutative space-time, denoted by $\kappa$-Minkowski space
\cite{kappa}. In this framework the Poincar\'{e} lie algebra itself
is deformed. Other methods lead to energy-momentum dependent
space-time metrics, called ``rainbow metrics" \cite{rain}, recently
investigated with the formalism of Finsler geometries \cite{fisler}.
Nevertheless, for the purpose of the present paper we are going to
make some simple physical considerations only in momentum space.

DSR, at least in its original guise, is formally independent of
gravity, even if the corrections it makes to SR are interpreted as
effective description of the imprints of quantum geometry in form of
some texture of flat space, present even in the limit of vanishing
gravi\-tational field. Thereby gravity is mainly needed as an
explanation for the departure of physical energy-momentum
conservation in DSR in such a way that the gravitational field is
thought as a reservoir for the non-conserved energy and momentum,
without specification of its properties at extremely small distances
and of the way it interacts with matter. A more concrete relation to
gravity exists in the Hopf algebra approach, as Hopf algebra methods
appear as a branch of quantum gravity research in their own right
\cite{maj}. Concerning the relation of DSR to Loop Quantum Gravity,
in \cite{abl} there is a rather heuristic derivation of modified
energy-momentum relations, inherent to DSR, from spatial
discreteness, but it is also explained that a violation of Lorentz
invariance can neither be derived nor excluded from the present form
of Loop Quantum Gravity.

Although DSR is supposed to reproduce gravitational effects in the
quantum gravity regime, it is an open issue, how it compares to
classical gravity. One may ask whether its low energy limit should
be in accordance with the effects of a classical gravitational
field, the low energy limit of quantum gravity.

In the literature there are essentially two different points of
view, relating DRS to different partial aspects of full General
Relativity (GR):
\begin{itemize}
\item No relation to classical gravity is proposed in \cite{top},
where DSR represents the topological degrees of freedom of the
gravitational field, i.\,e. the remnant, when the local degrees of
freedom of the gravitational field is removed. This approach is
supported by the successful formulation of 2+1 gravity, well known
to be a topological theory, as kind of DSR \cite{2+1}.
\item On the other
hand, in \cite{rain} a ``Correspondence Principle" is formulated in
the form that in the low-energy limit of DSR classical GR should be
recovered.
\end{itemize}

A logically different approach to the Lorentz invariance problem is
to separate between a particle's energy and momentum on the one hand
and the frequency and wave vector of the associated quantum wave
function on the other hand, with the advantage of an immediate
connection between the formulations in momentum and in position
space and a clear physical meaning of all variables in the game
\cite{SH}. An important consequence is an energy dependent Planck
constant, leading to modified uncertainty relations (Generalized
Uncertainty Principle, GUP) and possibly, but not necessarily, an
energy dependent speed of light.

The above-mentioned correspondence principle and the interpretation
of DSR as topological gravity being not equivalent, it is worthwhile
to compare low-energy approximations of DSR and GUP to classical
gravity in its simplest, i.\,e. Newtonian form. The program of the
present paper is to perform an elementary test, namely to apply
these approximations to a scattering process, as possible physical
effects always arise in connection with interactions between moving
objects. In the current DSR philosophy pseudo-variables must be
associated to interaction processes, so it is sufficient and
logically convenient to define them only in interaction regions, as
it was done in \cite{SH}, whereas the asymptotic variables are the
physical ones. The inclusion of interacting objects opens a door to
the introduction of gravity as a further interaction, leading to a
small perturbation, and not as a quantum property of space.

The only further ingredients, used beside Newtonian gravity in the
next sections, are the mass-energy relation and the de Broglie
wavelength of particles. In detail we will use the following
approximations:
\begin{itemize}
\item Newtonian gravity, understood as lowest-order approximation of
general relativity, in other words, as a simplified substitute for a
curved background.
\item Quantum field theory in first-order perturbative approximation.
\item A general lowest-order ansatz for DSR-like corrections of SR.
\end{itemize}
We are going to compare only unspecified interactions in the absence
and in the presence of classical gravity, so the considerations are
independent of specific high-energy quantum effects, like varying
coupling constants. In the next two sections, $c$ and $\hbar$ are
set equal to one, they will have to be restored in section 4.

\section{Gravity in two-particle interactions}
\subsection{Central collision}
We consider the scattering of two identical particles with repulsive
interaction in the centre-of-momentum reference frame. In
perturbative quantum field theory the free particles approach each
other, exchange virtual interaction particles and then move away
freely. During the free motion the gravitational interaction of
particles does not play a role, but we will take it into account in
the interaction process. If we assume first a central collision and
restrict ourselves to first-order Feynman diagrams, we can describe
the situation as follows. At the interaction vertices, when the
particles reach a certain minimal distance, they stop and their
kinetic energy materializes as a virtual exchange particle. Provided
the asymptotic kinetic energy is high enough, the gravitational
field of the virtual particle furnishes a significant amount of
additional energy for the interaction process in comparison with the
gravitation-less interaction, and the particles come closer to each
other, as if they had a higher asymptotic kinetic energy. In the
following we are going to formulate these considerations up to first
order in the gravitational constant $G$.

We assume two particles with masses $m$ and (absolute values of)
asymptotic momenta $p$. In the absence of gravity, at the
interaction vertices, with the particles at their minimal distance
denoted by $2r_0$, the asymptotic kinetic energy of both of them
transforms into the energy of the interaction particle,
\begin{equation}\label{Ek}
{\cal E}=2\left(\sqrt{p^2+m^2}-m\right).
\end{equation}
When Newtonian gravity is added to the system and the particles are
assumed to be massive, there are two effects (the mutual attraction
of the two rest masses is considered as negligible): Due to
gravitational attraction each particle has a potential energy
\begin{equation}
\Delta E_1=-\frac{G{\cal E}m}{r_0}
\end{equation}
in the moment when it stops at a distance $r_0$ from the scattering
centre. For a rough estimate of the minimal distance in terms of the
asymptotic kinetic energy we take the de Broglie wavelength
$\lambda$ of the exchanged particle, whose mass is assumed to be
negligible in comparison with its total energy, so that the
transmitted momentum is approximately equal to $\cal E$,
\begin{equation}\label{l}
2r_0\approx\lambda\approx\frac{1}{\cal E},
\end{equation}
and the gravitational potential energy of each of the scattered
particles becomes
\begin{equation}\label{E1}
\Delta E_1\approx-2\,\frac{m{\cal
E}^2}{m_P^2}=-8\,\frac{m\left(\sqrt{p^2+m^2}-m\right)^2}{m_P^2},
\end{equation}
where we have introduced the Planck mass $m_P=1/\sqrt{G}$, which, in
our units, stands also for the Planck energy and the Planck
momentum.

The second effect, which is independent of the mass of the
particles, is the self-energy $\Delta\cal E$ of the exchange
particle, whose order of magnitude is estimated by modeling it ad
hoc as a homogenous sphere of radius $r_0$,
\begin{equation}\label{cale}
\Delta{\cal E}\approx-\frac{3}{5}\frac{G{\cal
E}^2}{r_0}=-\frac{6}{5}\frac{{\cal E}^3}{m_P^2}.
\end{equation}
One half of $\Delta\cal E$ is associated to each of the scattered
particles to give rise to an energy difference
\begin{equation}
\Delta
E_2\approx-\frac{24}{5}\frac{\left(\sqrt{p^2+m^2}-m\right)^3}{m_P^2}.
\end{equation}
Of course, in view of our rough approximations and the homogenous
sphere being rather an indication of ignorance than a
seriously-meant model, the factors 6/5 and 24/5 appear ridiculous
and will be absorbed into order-of-magnitude factors later. For a
more exact description of the scattering of high-energy particles,
whose masses do not play a role, an Aichelburg-Sexl metric
\cite{Aic} would be convenient, for our considerations the above
simple estimate may be sufficient.

The interesting fact is that $\Delta E_1$ goes as $m{\cal E}^2$ and
$\Delta E_2$ as ${\cal E}^3$. As we are looking for gravitational
effects for realistic particles, we always have ${\cal E}\gg m$, so
that $\Delta E_1$, containing the rest mass $m$, will be normally
subdominant in comparison with $\Delta E_2$.

While the virtual particle, and in connection with it the
gravitational potential, come into being, the collision partners are
attracted and come closer to each other than they would in absence
of gravity. During this process the total energy is constant, the
kinetic energy increases and compensates the negative potential
energy. As it is only the kinetic energy, which plays a role in the
interaction process, we can replace gravity by an effective growth
of energy and momentum. On the other hand, after the collision the
particles are slightly slowed down by gravity and their asymptotic
outgoing energy is smaller than the energy immediately after the
interaction, so that asymptotic energy conservation is guaranteed.
(We do not assume graviton production, so that gravity is
conservative.)

So, instead of speaking about gravity, it is possible to ascribe an
effective energy $E_{\rm eff}$ to the particles, enlarged by
$-\Delta E_1$ and $-\Delta E_2$ in comparison with the asymptotic
values,
\begin{equation}\label{eps}
E_{\rm eff}=E-\Delta E_1-\Delta
E_2=E\left(1+\alpha\,\frac{mE}{m_P^2}+\beta\,\frac{E^2}{m_P^2}\right)
\end{equation}
with factors $\alpha$ and $\beta$ of the order of magnitude around 1
to 10.

Having ascribed an effective energy to the incoming particles, we
can also ascribe an effective momentum to them, simply by using the
free high-energy-momentum relation $E\approx p$,
\begin{equation}\label{pp}
p_{\rm
eff}=p\left(1+\alpha\,\frac{mp}{m_P^2}+\beta\,\frac{p^2}{m_P^2}\right).
\end{equation}
In some analogy to DSR, gravity is now hidden in effective
variables. (One might wonder whether a calculation involving a
Newtonian potential can be applicable to relativistic particles.
Relations (\ref{eps}) and (\ref{pp}) are justified by the fact that
the Newtonian potential is used only close to the turning points of
the particles, when they slow down to nonrelativistic velocities.)

Note that we have considered interactions in first order of a
perturbative expansion. In higher order, when one takes into account
more vertices, the interaction process becomes smoother, it is
divided into more steps and sets in earlier, i.\,e. at larger
distances, than in first order. In consequence, in higher order
diagrams the influence of gravity will become weaker, so the above
first-order estimates are rather an upper bound for gravitational
modifications.

To summarize, classical gravity influences the in- and outgoing
particles when they are close to their vertices, if the energy is
sufficiently high. This is described in two kinds of variables, both
of which have an immediate physical meaning: The effective ones,
$E_{\rm eff}$ and $p_{\rm eff}$, appearing at the vertices and
entering cross section calculations, and the asymptotic ones,
denoted by $E$ and $p$, playing the role of ``bare" variables in
connection with classical gravity.

\subsection{Non-central collision}
In the central collisions considered above the minimal distance of
colliding particles and the energy of a virtual particle have been
modified, quantities that are hardly accessible to direct measuring,
whereas the actual asymptotic energy/momentum are unaffected, so the
discussion is physically rather meaningless so far. The situation
improves in the case of non-central collisions of two particles with
impact parameter $b$. In first order perturbation theory this is
described in the following way: A particle moves straight ahead to
the point of minimal distance $r_0$ from the scattering centre, its
interaction vertex. There its radial momentum reverses by the
exchange of a virtual particle and it flies away along a straight
line at a scattering angle $\vartheta_0$ from its ingoing direction.

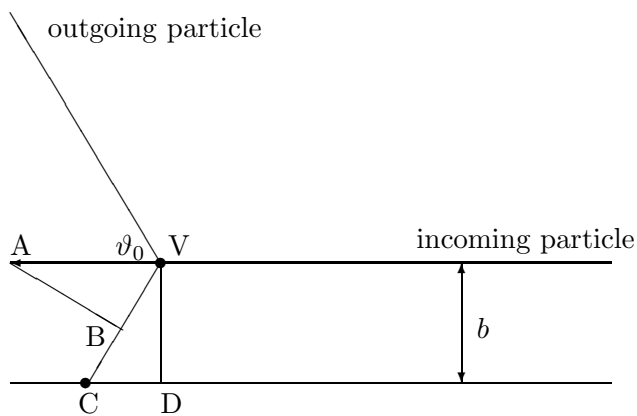
\setlength{\unitlength}{1mm}
\begin{picture}(100,65)(10,-15)
\put(10,10){\line(1,0){80}} \put(30,10){\vector(-1,0){20}}
\put(30,10){\circle*{1.5}} \put(30,10){\line(-3,5){20}}
\put(30,10){\line(-3,-5){10}} \put(10,10){\line(5,-3){15}}
\put(20,-6){\circle*{1.5}} \put(10,-6){\line(1,0){80}}
\put(30,10){\line(0,-1){16}} \put(31,11){V} \put(10,11){A}
\put(20,-1){B}\put(19,-10){C}\put(30,-10){D}\put(15,40){outgoing
particle}\put(64,12){incoming particle} \put(70,0){\vector(0,1){10}}
\put(70,0){\vector(0,-1){6}}\put(72,0){$b$}\put(24,11){$\vartheta_0$}
\end{picture}

In the figure the particle comes from the right, the scattering
centre is denoted by C and the vertex by V. A, B, and D are
auxiliary points. We may read off the following relations. The
triangles VCD and AVB are similar with the angles at $\angle$CVD and
$\angle$BAV being equal to $\vartheta_0/2$. The radial component
$\overrightarrow{\rm VB}$ of the momentum $\vec
p=\overrightarrow{\rm VA}$ at V is
\begin{equation}
p_r=p\sin\frac{\vartheta_0}{2},
\end{equation}
the relation between the particle's minimal distance
$r_0=\overline{\rm VC}$ from C and the impact parameter is
\begin{equation}
b=r_0\cos\frac{\vartheta_0}{2}.
\end{equation}

At the vertex $p_r$ becomes zero for a moment, so that during the
interaction process the kinetic energy is given only by the
component orthogonal to it, namely $\overline{\rm
BA}=p\cos\frac{\vartheta_0}{2}$. The energy, contributed from both
ingoing particles to the virtual exchange particle is therefore
equal to
\begin{equation}
{\cal
E}=2\left(\sqrt{p^2+m^2}-\sqrt{p^2\cos^2\frac{\vartheta_0}{2}+m^2}\,\right).
\end{equation}

Under the assumption $p\gg m$ the exchanged energy $\cal E$ can be
expanded in two different ways, according to the scattering angle
$\vartheta_0$. If $\vartheta_0$ is small, ${\cal
E}\approx2p(1-\cos\frac{\vartheta_0}{2})$ is small, too. The
particles do not slow down much and remain relativistic and the
considerations of the foregoing subsection, involving a Newtonian
potential, become inappropriate.

In the other case, when the collision is almost central and
$\vartheta_0$ is close enough to $180^\circ$, so that
$\cos\frac{\vartheta_0}{2}\ll\frac{m}{p}$, the particles slow down
to nonrelativistic speed and the calculations with Newtonian gravity
are more reliable. Now the energy transfer
\begin{equation}\label{et}
{\cal
E}\approx2p\left[1-\frac{m}{p}\left(1+\frac{1}{2}\frac{p^2}{m^2}\cos^2\frac{\vartheta}{2}
\right)\right]
\end{equation}
is large and for the wavelength associated with the exchange
particle we can again use the relativistic relation (\ref{l}),
$\lambda_0=1/\cal E,$ giving an estimate for the minimal distance
$2r_0$ of the particles.

An expansion of the potential energy $\Delta E_1$ of the rest masses
of the particles in the gravitational field of the virtual particle,
and the gravitational self-energy $\Delta\cal E$ of the latter one
yields
\begin{equation}
\Delta E_1\approx-2\,\frac{m\,{\cal E}^2}{m_P^2}\approx
-8\,\frac{mp^2}{m_P^2}\left[1-2\frac{m}{p}\left(1+\frac{1}{2}\frac{p^2}{m^2}
\cos^2\frac{\vartheta_0}{2}\right)\right]
\end{equation}
and
\begin{equation}
2\Delta E_2=\Delta{\cal E}\approx-\frac{6}{5}\frac{{\cal
E}^3}{m_P^2}\approx-\frac{48}{5}\frac{p^3}{m_P^2}\left[1-3\frac{m}{p}\left(1+\frac{1}{2}
\frac{p^2}{m^2}\cos^2\frac{\vartheta_0}{2}\right)\right].
\end{equation}

With the leading contributions of these corrections the effective
energy of the virtual particle, ${\cal E}_{\rm eff}={\cal E}-2\Delta
E_1-\Delta\cal E$, becomes
\begin{equation}\label{aff}
{\cal E}_{\rm eff}\approx
2p\left[1+\frac{24}{5}\frac{p^2}{m_P^2}-\frac{4}{5}\frac{mp}{m_P^2}\left(
4+9\frac{p^2}{m^2}\cos^2\frac{\vartheta_0}{2}\right)\right].
\end{equation}
The last two terms in parenthesis  are of the same order, because
$\cos^2\frac{\vartheta_0}{2}$ is of order $m^2/p^2$. Importantly,
there is a leading order correction, quadratic in $p/m_P$,
independent from the scattering angle, and a smaller one, of order
$mp/m_P^2$, depending on $\vartheta_0$.

When gravity is again replaced by introducing the effective energy
of the exchange particle, the wavelength of the latter one becomes
in leading order (coming from $\Delta E_2$)
\begin{equation}
\lambda\approx\frac{1}{{\cal E}_{\rm
eff}}\approx\frac{1}{2p\left(1+\beta\frac{p^2}{m_P^2}\right)},
\end{equation}
where, for convenience, the fancy numerical factor $24/5$ is again
replaced by $\beta$.

In the figure this means that the particle comes closer to the
centre C, the vertex V is shifted a small distance to the left, so
that the distance $\overline{\rm CV}$ becomes $\lambda/2$ instead of
$\lambda_0/2$ and the scattering angle becomes modified from
$\vartheta_0$ to $\vartheta$. From the relation
\begin{equation}\label{imp}
2b=\lambda_0\cos\frac{\vartheta_0}{2}=\lambda\cos\frac{\vartheta}{2}
\end{equation}
we obtain the modification of the scattering angle
\begin{equation}\label{teta}
\cos\frac{\vartheta}{2}\approx\left(1+\beta\,\frac{p^2}{m_P^2}\right)
\cos\frac{\vartheta_0}{2}.
\end{equation}
Due to the universality of gravity the ``bare" scattering angle
$\vartheta_0$ is unobservable, but it is possible to compare
(\ref{teta}) to the analogous result from DSR, obtained in the next
section.

\section{Comparison with DSR}
Now we are in a position to compare the two sets of variables
constructed in the foregoing section with the ``physical" and the
``pseudo"-variables in DSR. Once the deformed, nonlinear relations
for the physical variables are derived and modified kinematic
relations are established with the aid of the linear
pseudo-variables, they can in principle be forgotten, and all the
consequences, the deformed dispersion relations between energy and
momentum, the ensuing violations of conservation laws, etc. are
ascribed to gravity.

Here we go the opposite way by asking the question whether a
gravity-motivated deformation of SR is in its first approximation
compatible with DSR. In the foregoing section we have seen that
gravity influences particle scattering in the same way as if the
particles had a slightly higher effective kinetic energy. In the
following considerations this enhanced effective energy-momentum is
set into relation with the DSR pseudo-variables and the actual
asymptotic kinetic energy is related to the physical variables, as
usual \cite{ww}. Also in view of the desired parallel between DSR
and gravity this association of variables is plausible in the
following extrapolation: The unbounded pseudo-variables describe the
situation with a repulsive potential, that goes to infinity at zero
distance, in the absence of gravity: To reach smaller and smaller
distances from each other, the particles must have arbitrarily high
asymptotic energies. In most cases the same is true in the presence
of Newtonian gravity, but the asymptotic energy necessary to bring
particles close together, is lower. This actual energy is described
by the physical variables, which are smaller. Moreover, being
bounded, they predict distance zero at a finite asymptotic energy,
thus mimicking a gravitational collapse, when the exchange
particle's energy reaches the Planck region. By this fact DSR is
closer to GR than to Newtonian theory.

Here, of course, we are going to compare only the leading
corrections stemming from the inclusion of classical gravity, as
well as from DSR, both based on the ratio $p/m_P$ ($=E/m_P$ in our
assumption). Whereas in DSR the power of these ratio is a matter of
an ad hoc definition, classical gravity in three space dimensions
fixes the lowest order to be two, due to the simple fact that
$G=1/m_P^2$. This is in contrast to linear DSR corrections,
considered in \cite{KS}, for example.

For the comparison of classical gravity and DSR in non-central
scattering, considered in subsection (2.2), we assume a typical
lowest-order DSR relation (for different kinds of such
approximations, see \cite{exp}) between $p$ and $\pi$
\begin{equation}\label{beta}
\pi\approx p\left[1+\kappa\left(\frac{p}{m_P}\right)^n\right]
\end{equation}
with a constant $\kappa$ of order not too far from unity and some
positive integer power $n$. This kind of relation is in good
accordance with the leading term $\propto p^2$ in (\ref{pp}),
derived in connection with central collisions, if $n=2$.

Considering an almost central collision from a DSR point of view, we
replace $p$ by $\pi$, the variable related to interaction proceses,
in the wavelength of the virtual particle, so that
$\lambda\approx1/2\pi$. Then from (\ref{imp}) we obtain the modified
scattering angle,
\begin{equation}\label{thet}
\cos\frac{\vartheta}{2}\approx\left[1+\kappa\left(\frac{p}{m_P}\right)^n\right]
\cos\frac{\vartheta_0}{2}
\end{equation}
and from comparison of (\ref{thet}) with (\ref{teta}) it follows
that (at least in the considered scattering example) the lowest
order correction term of DSR can agree with Newtonian gravity, when
it is quadratic in the ratio $p/m_P$. Then only the constants
$\beta$ and $\kappa$ must be matched. The result is also a first
order approximation in the scattering angle around $180^\circ$. To
consider further $\vartheta$-dependent terms does not make much
sense in the scope of the Newtonian framework, because for faster
scattering processes there would be significant general relativistic
corrections. As Newtonian gravity is the lowest order correction of
SR coming from GR, we have obtained a condition for DSR theories to
satisfy the correspondence principle in its full meaning, namely,
the lowest-order effects of DSR must be quadratic. DSR 2 for
example, proposed in \cite{MS}, with linear corrections would be at
odds with it.

One important difference of the present approach to ``full" DSR is
the use of the free energy-momentum relations for both kinds of
variables, rather than of modified ones for $E$ and $p$. This
important aspect of DSR does not show up in the present
calculations, because in the considered approximations the mass term
does not play a role, and $E\approx p$ as well as
$\varepsilon\approx\pi$, the calculations were essentially carried
out for momenta alone.

The above considerations can easily be applied to higher dimensions.
In $d>3$ space dimensions the Planck mass $m_P^{(d)}$ is by orders
of magnitude smaller, on the other hand, the Newtonian potential
goes as $r^{2-d}$. See for example \cite{G}. In consequence, the
lowest order correction of the scattering angle behaves as
\begin{equation}
\frac{p^{d-1}}{\left(m_P^{(d)}\right)^{d-1}},
\end{equation}
if there are compactified dimensions, large enough for classical
gravity to be a reasonable approximation when the minimal distance
is as small as the magnitude of these dimensions. In consequence, in
these cases the lowest-order corrections of DSR must be of order
$d-1$, when compatibility with classical gravity is desired.

\section{Comparison with GUP}
This approach is characterized by making a principal distinction
between the energy and momentum $(E,\vec p\,)$ of a particle and its
associated frequency and wave vector $(\omega,\vec k\,)$. Their
relation is most generally written as
\begin{equation}\label{Efp}
(\omega,\vec k\,)=\left(E\cdot f(E,\vec p\,),\vec p\cdot g(E,\vec
p\,)\right).
\end{equation}
Energy and momentum are assumed to be unbounded, whereas $\omega$
and $\vec k$ are bounded by orders of magnitude 1/{\sl Planck time}
and 1/{\sl Planck length}, respectively. The functions $f$ and $g$
can be chosen analogously to the functions relating physical and
pseudo-variables in arbitrary versions of DSR. Nevertheless, the
interpretation is different: There are no merely auxiliary
variables, both $(E,\vec p\,)$ and $(\omega,\vec k\,)$ have a clear
physical meaning and there is a natural relation between momentum
and position space from the beginning.

Comparing (\ref{Efp}) with the standard relation
\begin{equation}
(E,\vec p\,)=\hbar(\omega,\vec k\,),
\end{equation}
one finds energy-momentum dependent constants
\begin{equation}
\tilde{\hbar}(E,\vec p\,)=\frac{1}{f(E,\vec p\,)}
\hspace{2cm}\mbox{and}\hspace{2cm} \tilde{c}(E,\vec
p\,)=\frac{\omega}{|\vec k|}=c\,\frac{f(E,\vec p\,)}{g(E,\vec p\,)}.
\end{equation}
For high energies $\tilde\hbar$ increases, increasing the
quantum-mechanical uncertainties. We shall restrict ourselves to the
case $\tilde c=c$, i.\,e. $f=g$.

In analogy to (\ref{beta}) we assume a lowest-order relation
($p=|\vec p|$ and $k=|\vec k|$)
\begin{equation}
p=\tilde{\hbar}k\approx\hbar
k\left[1+\kappa\left(\frac{p}{m_P}\right)^n\right],
\end{equation}
leading to the wavelength
\begin{equation}
\lambda=\frac{1}{k}\approx\frac{\tilde\hbar}{2p}.
\end{equation}
(Note that $k$ is the wave vector of the exchange particle, $p$ is
the momentum of one incoming particle.) As before, from (\ref{imp})
one obtains the correction of the scattering angle,
\begin{equation}
\cos\frac{\vartheta}{2}\approx\left[1-\kappa\left(\frac{p}{m_P}\right)^n\right]\cos\frac{\vartheta_0}{2}.
\end{equation}
With the same choice of transformation functions between the
different sets of variables in first approximation, GUP has yielded
just the opposite sign of the DSR correction in (\ref{thet}).

\section{Conclusion}
One main result of the considerations of this paper is the condition
that DSR corrections to SR must be quadratic in $p/m_P$ in lowest
order to fulfill the correspondence principle in the given example.
It is not shown that this condition is sufficient in every situation
and the calculations do not show how DSR differs from GR, when
higher particle energies are involved. A relation of the present
result with the interpretation of DSR in \cite{top} as the
topological part of GR would depend on the properties of particle
trajectories in topological 3+1 gravity.

The second result concerns GUP, where the situation is quite
different. DSR produces an effective attractive force, GUP, on the
other hand, results in a repulsive force, which is not a big
surprise, as it lays lower bounds to space and time intervals. In
contrast to DSR, rather than competing with Newtonian gravity (and
thus GR), GUP counteracts it, thus qualifying as a description of
pure quantum gravity effects, which has nothing at all to do with
classical gravity. In order not to collide with GR, the lowest-order
term in GUP must be of a higher power than 2. There is an
interesting parallel to loop quantum cosmology \cite{LQC}, where a
repulsive behaviour of gravity at short distances, which helps to
avoid singularities, is observed.

In the considered example the effects of DSR and those of GUP would
be equivalent for some $n>2$, if the roles of $(E,\vec p\,)$ and
$(\varepsilon,\vec\pi)$ were interchanged. $\varepsilon$ and
$\vec\pi$ would be energy and momentum ascribed to free particles,
which can be boosted to arbitrary values with respect to a certain
reference frame, as long as no interaction takes place. The physical
energy $E$ and momentum $\vec p$, on the other hand, play a role in
interactions, which would be in accordance with their interpretation
as measurable quantities, as measurements always go along with some
interactions. Due to the exchanged roles of $(E,\vec p\,)$ and
$(\varepsilon,\vec\pi)$ in relation to the common DSR
interpretation, reaction thresholds anomalies would be equally small
as in usual DSR, but in the opposite direction. For example, when
conventional DSR predicts an insignificant lowering of the GZK
cutoff \cite{SM}, the reversed interpretation would lead to an
(equally insignificant) raising. (Recent observations do not seem to
confirm a shift of the GZK cutoff at all \cite{GZK}.)\\

{\bf Acknowledgement}. Supported by the Ministry of Education of the
Czech republic, contract number MSM 0021622409. The author thanks
the Perimeter Institute for warm hospitality and support and S.
Hossenfelder and K. Bering for helpful discussions.

\end{document}